\documentclass[aps,prl,preprint,longbibliography]{revtex4-1}
\usepackage{amsmath}
\usepackage{graphicx}
\usepackage{pstricks}
\usepackage{pst-node}
\usepackage{pst-coil}
\usepackage{pst-grad}
\usepackage{multido}
\usepackage{dcolumn}
\usepackage{bm}
\usepackage[normalem]{ulem}

\newcommand{{\invcm}}{\ensuremath{\,\mathrm{cm}^{-1}}}

\begin{document}

\title{Stokes-anti-Stokes correlated photon properties akin to photonic Cooper pairs}

\author{Filomeno S. de Aguiar J\'unior$^1$}
\author{Andr\'e Saraiva$^2$}
\author{Marcelo F. Santos$^2$}
\author{Belita Koiller$^2$}
\author{Reinaldo de Melo e Souza$^3$}
\author{Arthur Patroc\'\i nio Pena$^1$}
\author{Raigna A. Silva$^{1,4}$}
\author{Carlos H. Monken$^1$}
\author{Ado Jorio$^1$}
\affiliation{$^1$Departamento de F\'\i sica, UFMG, Belo Horizonte, MG 31270-901, Brazil }
\affiliation{$^2$Instituto de F\'\i sica, UFRJ, CP 68528, Rio de Janeiro, RJ 21941-972, Brazil}
\affiliation{$^3$Instituto de F\'\i sica, UFF, Niter\'oi, RJ 24210-346, Brazil}
\affiliation{$^4$Instituto de F\'\i sica, UFU, Uberl\^andia, MG, Brazil}

\date{\today}

\begin{abstract}
Photons interact with each other in condensed matter through the same mechanism that forms Cooper pairs in superconductors -- the exchange of virtual phonons [PRL 119, 193603 (2017)]. It is however unclear which consequences of this interaction will be observable and potentially lead to further analogy with superconductivity. We investigate the energy, momentum and production rate of correlate Stokes-anti-Stokes ({\it SaS}) photons in diamond and other transparent media, experiencing properties akin to those of electronic Cooper pairs. The rate of correlated {\it SaS} production depends on the energy shifts of the pair, which in the BCS theory determines whether there should be an attractive or repulsive interaction. With this view, we only observe correlated {\it SaS} in the case of attractive interactions. While traditional photon-phonon collisions scatter light in all directions, the correlated {\it SaS} photons follow the same path as the noninteracting laser. The observed correlated {\it SaS} photon pairs are rare, but our model indicates paths to achieve higher interaction energies.
\end{abstract}

\maketitle

The production of red-shifted (Stokes) and blue-shifted (anti-Stokes) photons by inelastic scattering of light in matter, where the incoming laser photons of energy $E_L = \hbar \omega_L$ may lose or gain energy in the form of atomic vibrations (phonons) of energy $E_q$, is known as Raman scattering  \cite{RAMAN1928ARadiation,Walls1970QuantumEffect} and it is used for characterizing materials properties in materials science studies \cite{kuzmany2009solid}. By selecting detection events that happen within a short time interval (fs to ps range) \cite{del2012theory} and symmetrically shifted in frequency from the excitation laser mode, we are able to identify correlated {Stokes-anti-Stokes} (\textit{SaS}) photon pairs \cite{Kasperczyk2016TemporalWater}. They come from events in which the same phonon created in the sample by the Stokes (\textit{S}) process is annihilated by the anti-Stokes (\textit{aS}) process \cite{klyshko1977correlation,Parra-Murillo2016Stokesanti-StokesFunction}. Several recent studies \cite{lee2011entangling,lee2012macroscopic,england2013photons,riedinger2016non,anderson2018two,Kasperczyk2016TemporalWater} explored the production of \textit{SaS} pairs through real processes, \textit{i.e.} when the energy (Raman) shifts $\varepsilon_{aS}$ and $\varepsilon_{S}$ correspond respectively to plus and minus a quantum of vibration $E_{q}$ in the material (resonant process). Their main motivation is the potential applications of \textit{SaS} pairs in quantum information.

The production of \textit{SaS} photon pairs may occur out of resonance ($|\varepsilon_{S,aS}| \neq E_{q}$, $\varepsilon_{S} = -\varepsilon_{aS}$ for energy conservation), in a process we call virtual \textit{SaS}, viewed as the photonic counterparts of superconducting Cooper pairs \cite{Shen2002PCP,Saraiva2017PhotonicPairs,zhang2018generation}. An analogy between the virtual \textit{SaS} and photonic Cooper pairs (PCPs) was then proposed \cite{Saraiva2017PhotonicPairs}, but there is so far no 
exploration of the properties akin to those of PCPs \cite{Shen2002PCP} and photonic four-wave mixing \cite{Fan2009,Takesue2012,Caspani2017}.

In the second quantization, any two particle interaction Hamiltonian can be described in the form \cite{ballentine2014quantum}
\begin{equation}
\hat{H}_{int}= \sum_{\mathbf{k_1},\mathbf{k_2},\mathbf{k_3},\mathbf{k_4}} V (\mathbf{k_1},\mathbf{k_2},\mathbf{k_3},\mathbf{k_4}) \,\hat{a}_{\mathbf{k_4}}^\dagger \hat{a}_{\mathbf{k_3}}^\dagger \hat{a}_\mathbf{k_2} \hat{a}_\mathbf{k_1},
\label{eq:hamiltonian}
\end{equation}
where ${\rm \bf k_i}$ labels the quantum states. This $\hat{H}_{int}$ can be used to describe electron-electron coupling in superconductivity, non-linear photon-photon processes, and any two-particle interaction phenomenon, with the specificities residing in the interaction potential $V (\mathbf{k_1},\mathbf{k_2},\mathbf{k_3},\mathbf{k_4})$. All such processes represent four-wave mixing, although this terminology is generally used only in the field of optics \cite{Boyd2003NonlinearOptics}, where $V (\mathbf{k_1},\mathbf{k_2},\mathbf{k_3},\mathbf{k_4})$ is associated with a third-order electrical susceptibility.

\begin{figure}
\includegraphics[width=8cm]{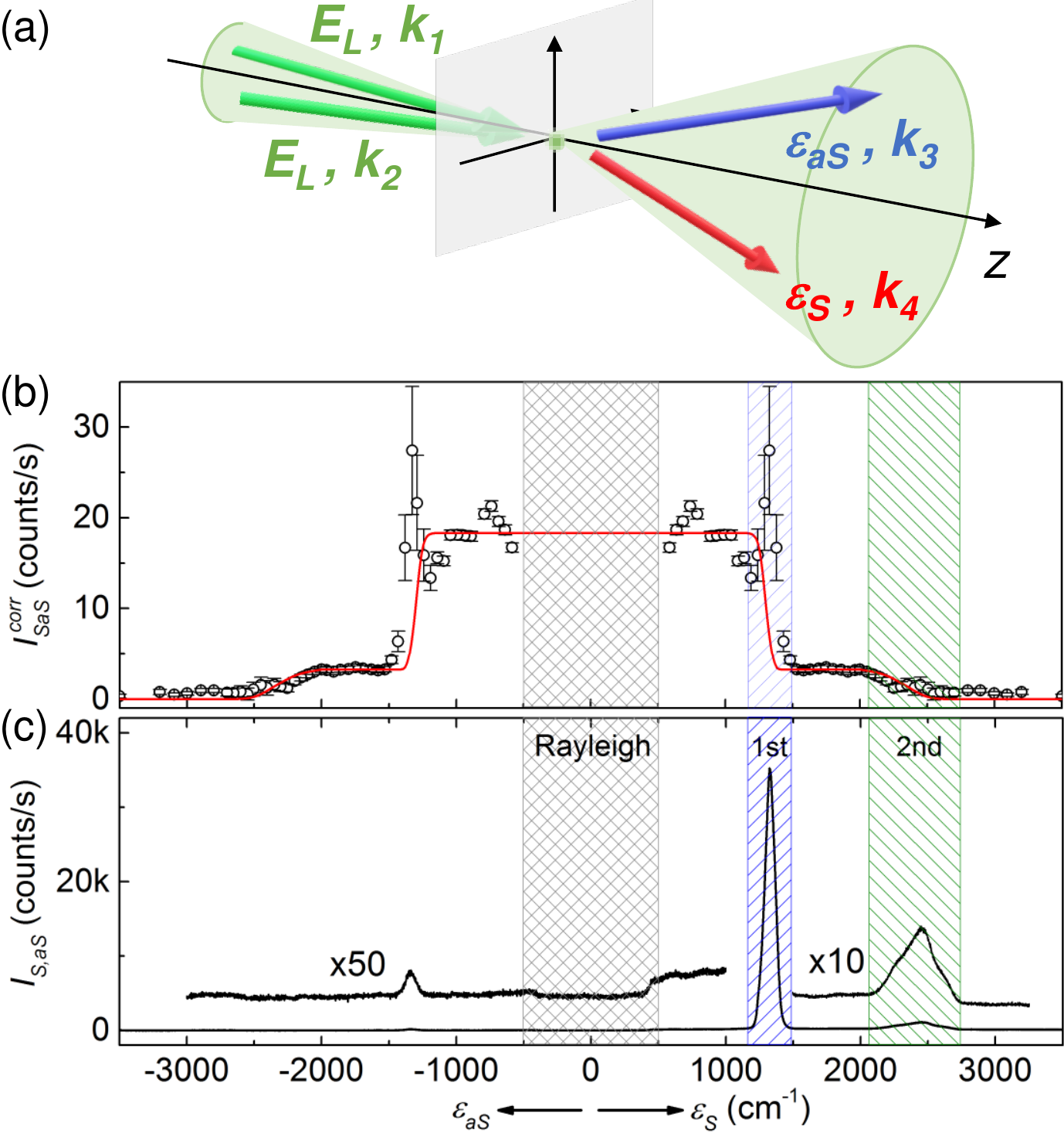}
\caption{(a) Schematics showing light-by-light scattering. (b) The black circles span the number of correlated {\it SaS} measured per second, under the excitation power of $P_L = 40$\,mW. The error bars are taken as the sum of $sqrt(N)$ (per second) for $I_{SaS}(\Delta t = 0)$ and $\overline{I}_{SaS}(\Delta t \neq 0)$, where $N$ is the total number of events observed during 600s accumulation time per data point. The red line is a fitting to the data considering Eq.\,\ref{eq:SaSepsilon}. (c) The black solid line gives the Raman spectrum.}
\label{figure1}
\end{figure}

A billiard-like picture representing such a photon-photon interaction is depicted in Figure~\ref{figure1}(a). This process is implemented experimentally with the incoming laser beam focused inside a diamond slab of 1.7 mm by a microscope objective of low numerical aperture (NA = 0.6), and the forward scattered light collimated by another microscope objective of high NA = 0.9 in a confocal arrangement \cite{suppinfo}. The sample is excited with a $T_L = 200$\,fs width pulsed laser at $R_L = 76$\,MHz pulse rate, wavelength $\lambda_L = 633$\,nm, and the PCPs are selected by time filtering only \textit{S} and \textit{aS} fotons that arrive in two different photon counters (avalanche photodiodes, APDs) in the same laser pulse (time delay $\Delta t = 0$ \cite{suppinfo}). Accidental coincidences also happen (uncorrelated \textit{S} and \textit{aS} fotons measured at $\Delta t = 0$) \cite{kasperczyk2015stokes}, and they can be filtered considering the correlated {\it SaS} count rate given by
\begin{equation}
I_{SaS}^{corr} = I_{SaS}(\Delta t = 0) - \overline{I}_{SaS}(\Delta t \neq 0)\, ,
\label{eq:PCP}
\end{equation}
where the overline indicates average over the measured \textit{SaS} count rate $I_{SaS}(\Delta t \neq 0)$, valid because for coherent fields the normalized second-order correlation function 
$g^2(\Delta t = 0) = I_{SaS}(\Delta t = 0)/\overline{I}_{SaS}(\Delta t \neq 0) = 1$.

Figure~\ref{figure1}(b) shows $I_{SaS}^{corr}$ (black circles) for different values of $\varepsilon_S$ and $\varepsilon_{aS}$. A single experimental detection measures the number of \textit{S} and \textit{aS} photon pairs reaching the two APDs at the same time ($\Delta t = 0$), counting during 600 s, and the same data point is represented twice, in both $\varepsilon_{S}$ and $\varepsilon_{aS}$ sides of the graphic. Only the \textit{S} beam is spectrally filtered using a monochromator (26\invcm resolution) to simplify the spectral filtering dependence, since we have already established that the correlated {\it SaS} only exists for $\varepsilon_{S} = -\varepsilon_{aS}$ \cite{Kasperczyk2016TemporalWater,Saraiva2017PhotonicPairs}. The accidental coincidences depend whether the \textit{aS} beam is filtered or not, but this is irrelevant for the correlated {\it SaS} counting.

In Figure~\ref{figure1}(c) the black line gives the Raman intensity $I_{S,aS}(\varepsilon_{S,aS})$ of the sample, measured with a spectrometer equipped with a charge coupled device (CCD). The result is quantitatively consistent with the $I_{S,aS}(\varepsilon_{S,aS})$ measured with one APD replacing the CCD and using the spectrometer as a monochromator. We adopt the usual convention in Raman spectroscopy, representing the Stokes shift in the plot as positive [$\varepsilon_S = -(E_{L}-E_{S})$, while $\varepsilon_{aS} = -(E_{L}-E_{aS}) <0 $]. The black-hatched area indicates the Rayleigh spectral region, removed with a notch filter, and the blue- and green-hatched areas in the Stokes side indicate the ranges of
$1^{st}$-order and $2^{nd}$-order Raman spectral responses, respectively \cite{suppinfo}.

From panels (b) and (c) in Fig.~\ref{figure1} we conclude that  $I_{SaS}^{corr}(\varepsilon_{S,aS})$ is highest for pairs formed by virtual phonons with $|\varepsilon_{S,aS}| < E_{q=0} = 1332${\invcm} \cite{solin1970raman}, dropping significantly once this first-order Raman peak is crossed. Correlated {\it SaS} are also observed, with lower count rates, between $E_{q=0}$ and the second-order (two-phonon scattering, with $+q$ and $-q$ non-zero momenta) Raman feature at $2E_{q} = 2500${\invcm}, which comes from a peak at $E_{q \neq 0} \sim 1250${\invcm} in the diamond phonon density of states \cite{solin1970raman}, and it drops again when crossing the second-order Raman peak.

The energy behavior in the correlated {\it SaS} efficiency can be explained using Eq.\,\ref{eq:hamiltonian} to investigate the quantum state $|\psi_f\rangle$ of the outgoing correlated {\it SaS}, where {\bf k$_{\rm i}$} labels the four photon momenta (see Fig.\,\ref{figure1}(a)), (${\rm{\bf i}}=1,2$) for the incident (laser) photons and (${\rm{\bf i}}=3,4$) for the scattered (\textit{aS} and \textit{S}) photons, and $\hat{a}_\mathbf{k_i}$ are photon annihilation operators \cite{Saraiva2017PhotonicPairs}. $|\psi_f\rangle = e^{-iH_{int}t}|\psi_0\rangle \sim (\mathbf{1}+i\hat{H}_{int}dt/\hbar)|\psi_0\rangle$, with $|\psi_0\rangle = |\alpha_{L}\rangle |00\rangle$, where $|\alpha_{L}\rangle$ represents the coherent laser state, $|00\rangle$ the \textit{S} and \textit{aS} vacuum state. The correlated {\it SaS} are produced mainly with the same polarization of the incident laser \cite{Kasperczyk2016TemporalWater}, however, for simplicity, we do not consider polarization here.

The $I_{SaS}^{corr}(\varepsilon_{S,aS})$ is due to spontaneous Raman scattering, driven by the vacuum of phonon, \textit{S} and \textit{aS} photon fields. This is the case because in the maximum (resonant) observed value of $I_{S}(\varepsilon_S=1332{\invcm}) \sim 35$\,kilocounts/s (see Fig.\,\ref{figure1}(c)), the probability to generate a Stokes photon in one pulse is $10^{-4}$, and much less for \textit{aS}. The phonon lifetimes (fs to ps range) are much shorter than the time-distance between pulses (13\,ns), so that the correlated {\it SaS} production happens necessarily within one pulse, which is with a very high probability in the vacuum state of \textit{S} photons, of \textit{aS} photons and of phonons (for diamond $E_q$ is much higher than the room temperature thermal energy). 

The most important aspect in Fig.\,\ref{figure1} is the roughly constant correlated {\it SaS} rate in energy, but highly asymmetric with respect to the resonant processes, which take place at the Raman-active phonon energies $E_{q=0}$ and $2E_{q \neq 0}$ in the $1^{st}$- and $2^{nd}$-order scattering processes, respectively. The energy dependence of the perturbative photon-photon coupling $V(\mathbf{k_1},\mathbf{k_2},\mathbf{k_3},\mathbf{k_4})$, as obtained in Ref.\cite{Saraiva2017PhotonicPairs}, describes a correlated {\it SaS} production rate that is symmetric with respect to the phonon energy $E_q$, and it does not fit the data. Other possibilities, such as losses (e.g. phonon decay), resonant and non-resonant Raman contributions, or quantum interference between the $1^{st}$- and $2^{nd}$-order Raman processes have also been considered, but they are not able to fit the data due to the relatively large asymmetry of the correlated {\it SaS} production rate above and below the Raman peak together with the relatively sharp (in width) and symmetric Raman peak.  Therefore, within the perturbative quantum mechanics framework introduced in Ref.\,\cite{Saraiva2017PhotonicPairs} the $I_{SaS}^{corr}$ dependence on $|\varepsilon_{S,aS}|$ is inexplicable. Notice that the results and  consequences of the BCS theory of superconductivity cannot be obtained within a perturbation theory framework based on unpaired unperturbed electrons, even if summed over all orders.

Akin to the BCS original theory \cite{Bardeen1957TheorySuperconductivity}, we adopt here the simplified description of the interaction potential
\begin{equation}
V (\mathbf{k_1},\mathbf{k_2},\mathbf{k_3},\mathbf{k_4}) = \left\{\begin{array}{cc}
	-V_0, &|\varepsilon_{S,aS}| < E_q, \\ 0, &|\varepsilon_{S,aS}| > E_q, \end{array}\right.
\label{eq:potential}
\end{equation}
i.e. a negative constant coupling between two photons when their \textit{SaS} Raman shift modulus is less than the energy of a real phonon, and zero elsewhere \cite{madelung2012introduction}.
Thus, $V(\mathbf{k_1},\mathbf{k_2},\mathbf{k_3},\mathbf{k_4})$ of Eq.\,\ref{eq:potential} implies that correlated {\it SaS} are formed in the attractive interaction range. The virtual particle mediating this interaction exists only during the very short time interval ($\lesssim$10\,fs) in which the photons coexist inside the $\sim 3$\,$\mu$m focal region of the pump laser beam, in a genuine photon-photon collision conserving energy and momentum. As for the familiar BCS Cooper pairs, photons deviated by energies corresponding to positive values of $V(\mathbf{k_1},\mathbf{k_2},\mathbf{k_3},\mathbf{k_4})$ interact repulsively and we empirically conclude that they do not form correlated {\it SaS}.
We may write the Raman shift dependence of the correlated {\it SaS} production rate as \cite{suppinfo}
\begin{equation}
I_{SaS}^{corr} = \Delta k |\alpha_L^2 V_0 \frac{T_L}{\hbar}|^2 R_L\,,
\label{eq:SaSepsilon}
\end{equation}
for $|\varepsilon_{S,aS}| < E_q$, where $|\alpha_{L}|^2$ is the number of pump laser photons per pulse ($1.8\times 10^{9}$ for $P_L = 40$\,mW), $\Delta k$ is the spectral collection obtained experimentally from the ratio between the monochromator resolution and the total scattering range of non-zero potential (1332{\invcm} for $1^{st}$-order and 2500{\invcm} for $2^{nd}$-order). Solid angle is not considered here because, as shown later, the correlated {\it SaS} cross the sample without momentum scattering. Since the interaction is mediated by phonons, the value of $V_0$ is proportional to the electron-phonon scattering efficiency squared $M_q^2$ \cite{Saraiva2017PhotonicPairs}, and it can be obtained directly from the Raman scattering intensity $I_S \propto M_q^2$, then $|V_0^{1st,2nd}| = C^{1st,2nd}A_S^{1st,2nd}$, where $A_S^{1st,2nd}$ is the area below the Stokes $1^{st}$- and $2^{nd}$-order Raman peaks, obtained experimentally from Fig.\,\ref{figure1}(c). $I_{SaS}^{corr}(\varepsilon_{S,aS})$ according to Eq.\,\ref{eq:SaSepsilon} is shown by the red solid line in Fig.~\ref{figure1}(b), with the fitting parameters $C^{1st}=5.75\times 10^{-22}$ and $C^{2nd}=3.35\times 10^{-21}$, in units of [\,eV$\cdot$cm$\cdot$s], adjusting the intensity levels below and above 1332{\invcm}.

\begin{figure}
\includegraphics[width= 8cm]{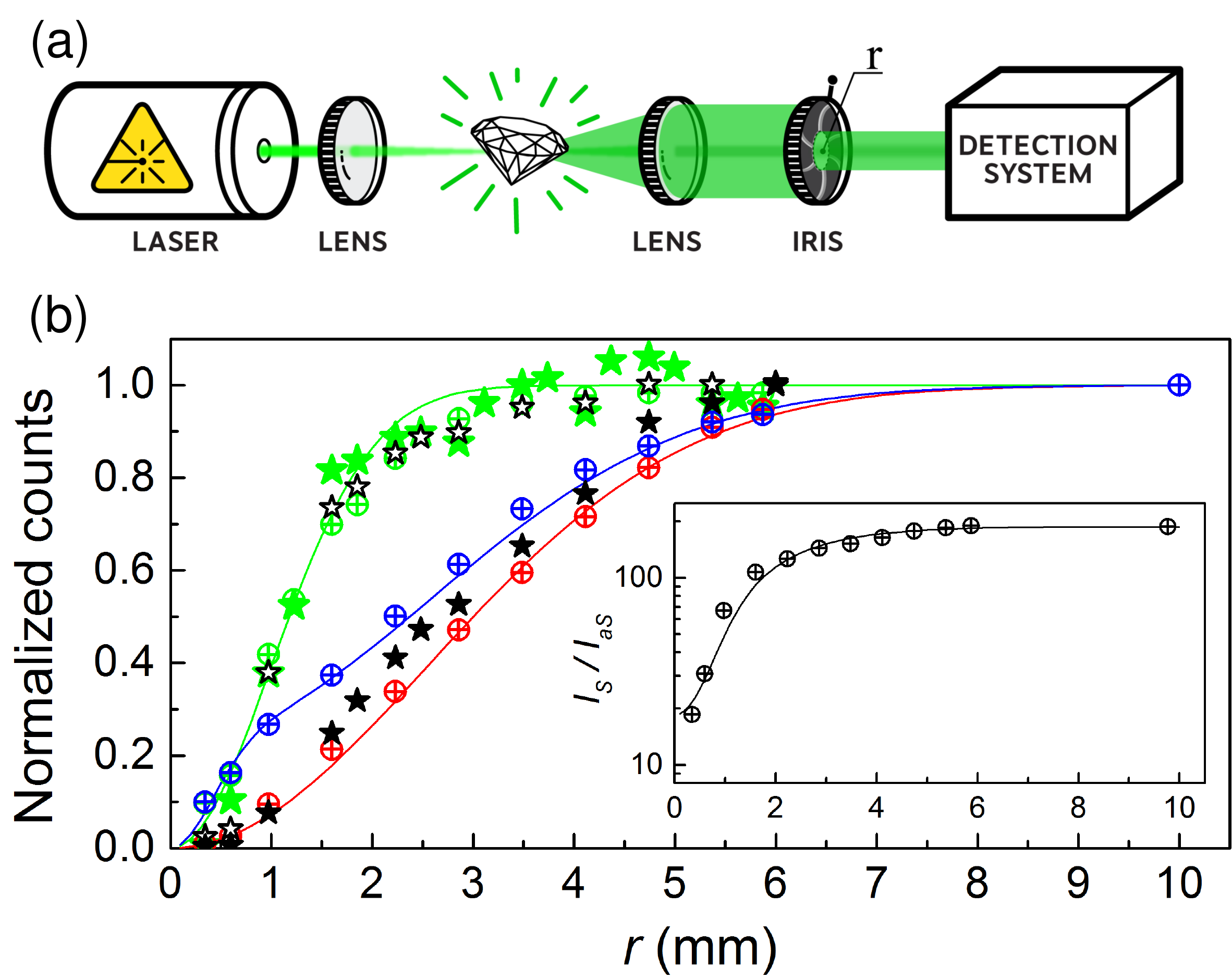}
\caption{(a) Angular spread for inelastic scattered light emerging after crossing a diamond slab, analyzed by re-collimating the scattered rays with a confocal lens and selectively blocking the outer rays with a variable radius aperture $r$ (iris). (b) Normalized iris aperture dependence for: the unpaired \textit{S} intensities (red crossed-circles); unpaired \textit{aS} intensities (blue crossed-circles); non-interacting excitation laser (green crossed-circles); time correlated \textit{SaS} photon pairs (green and black stars). The unpaired \textit{S} and \textit{aS} signals are collected at the real Raman peak energies ($\varepsilon_{S,aS} = \pm 1332{\invcm}$). The \textit{SaS} correlated photons are collected both at the Raman peak energy ($\varepsilon_{S,aS} = \pm 1332{\invcm}$, open and filled black stars) and outside ($\varepsilon_{S,aS} = \pm 900{\invcm}$, green stars), to select \textit{SaS} pairs created by real and virtual phonons, respectively. For the real \textit{SaS} we plot separately the true coincidences (correlated {\it SaS}, open symbols) and the accidental coincidences (filled symbols). The total counts $I_{SaS}(\Delta t = 0)$ (not shown) falls in between the two. The inset plots the ratio between the unpaired \textit{S} and \textit{aS} intensities measured at $\varepsilon_{S,aS} = \pm 1332{\invcm}$. Solid lines are fitting to the data.}
\label{figure2}
\end{figure}

Another interesting property of the emerging correlated {\it SaS} is given by momentum conservation (or photonic phase matching), where the billiard-like physics resulting from this interaction may be probed analyzing the transverse spatial correlation of the pairs, as depicted in Figure~\ref{figure2}. The angular spread of the scattered photons is analyzed by limiting the solid angle collected by the detection system with the help of a circular aperture (iris) of variable radius $r$, as shown schematically in Fig.\,\ref{figure2}(a).

Typically, photons ricochet in all directions when they scatter against phonons, resulting in an intensity profile for the Raman effect with a deviation from the forward propagation direction of the incident laser beam \cite{Schlosser2013EvaluationAberrations}. This is evidenced by the steady growth of the count rate of scattered \textit{aS} and \textit{S} photons as a function of the iris aperture shown by the red and blue crossed-circles respectively in Fig.\,\ref{figure2}(b).

In contrast, the non-resonant correlated {\it SaS} count (green stars in Fig.\,\ref{figure2}(b)) inherits the same spatial profile defined by the excitation laser (green crossed-circles in Fig.\,\ref{figure2}(b)), dropping significantly only when the iris is closed below $r = 2$\,mm. For resonant \textit{SaS}, where accidental coincidences are significant, $I_{SaS}^{corr}$ (open black stars) follows the laser dependence, while accidental coincidences (filled black stars) follow the unpaired \textit{aS} and \textit{S} photons tendency. Therefore, although the accidental coincidences are correlated in time, they belong to uncorrelated scattering processes, in other words, they are not correlated {\it SaS}. The correlated {\it SaS} cross the material following the same path as the noninteracting incident laser -- a phenomenon analogous to the transfer of amplitude profile in spontaneous parametric down conversion (SPDC)~\cite{suppinfo,Burnham1970ObservationPairs,boyer2008entangled,Walborn2010SpatialDown-conversion}, a hint for establishing photonic supercurrent behavior.

The data in Fig.\,\ref{figure2}(b) can be fitted considering a Gaussian distribution of the scattered intensities (solid lines \cite{suppinfo}). Regarding the real \textit{aS} data (blue crossed-circles in Fig.\,\ref{figure2}(b)), good fits are obtained considering a sum of two Gaussian distributions. This phenomenon is better visualized considering the intensity ratio $I_{S}/I_{aS}$ between the unpaired \textit{S} and \textit{aS} signals, shown in the inset to Fig.\,\ref{figure2}(b). This ratio  provides a figure of merit for both the thermally and the correlated {\it SaS} generated \textit{aS} signals playing a role in the observed scattering \cite{Parra-Murillo2016Stokesanti-StokesFunction}. The significant decay in $I_{S}/I_{aS}$ for $r < 2$\,mm demonstrates the \textit{aS} Raman signal is dominated by the correlated {\it SaS} in the low scattering angle region, providing a spatial-filter technique to reject uncorrelated \textit{S} and \textit{aS} signals.

\begin{figure}
\includegraphics[width= 8cm]{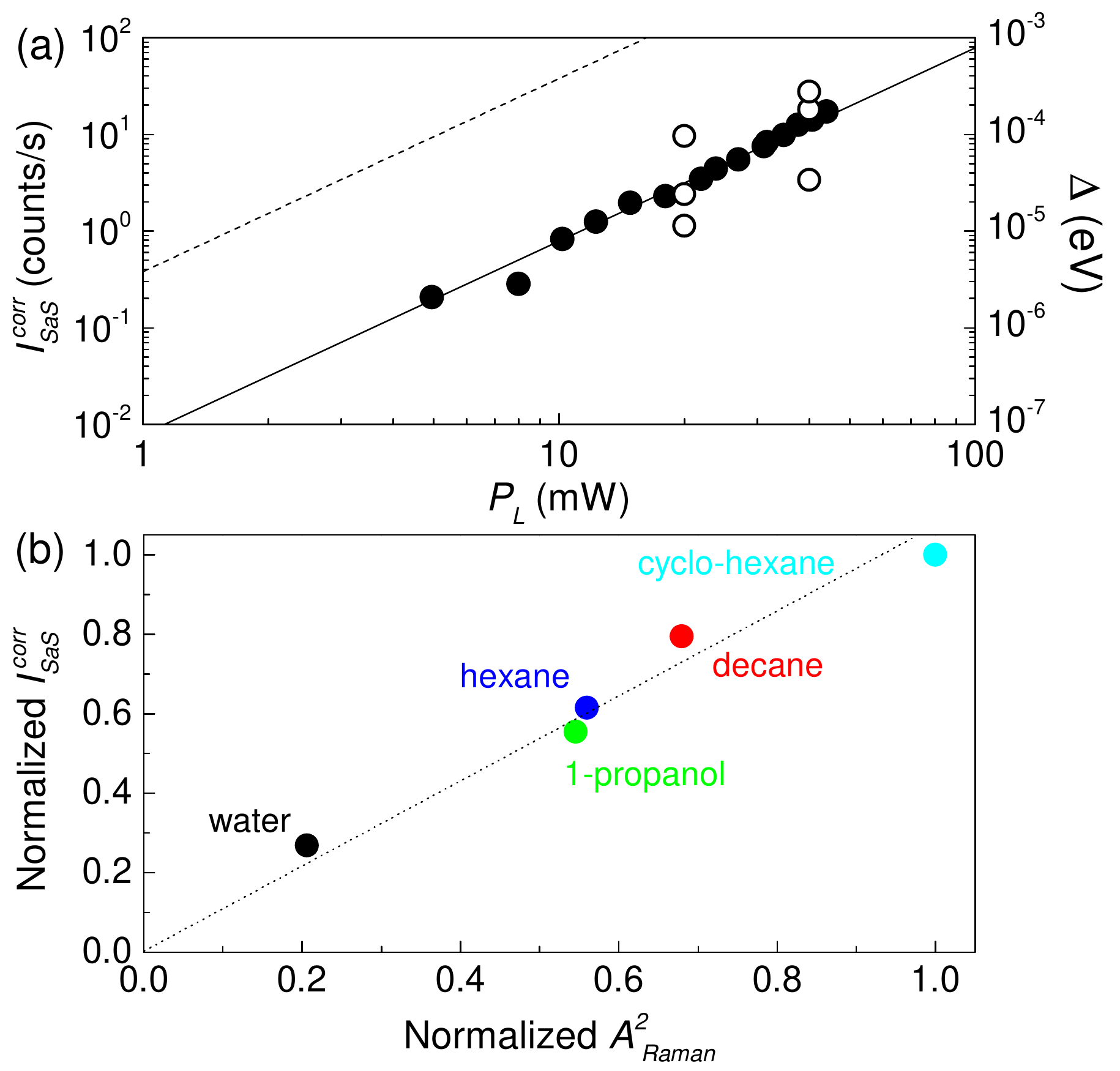}
\caption{(a) Correlated {\it SaS} rate $I_{SaS}^{corr}$ as a function of excitation laser power ($P_L$). The black circles following the solid line are measurements for diamond with a Raman shift of $\varepsilon_{S,aS}= \pm$900{\invcm}. For $P_L = 20$\,mW and 40\,mW (see open circles), $I_{SaS}^{corr}$ are also obtained at three different Raman shifts - namely $\pm$1700{\invcm} (above $E_{q=0}$), $\pm$900{\invcm} (below $E_{q=0}$) and $\pm$1332{\invcm} (at $E_{q=0}$), from lower to higher $I_{SaS}^{corr}$ values, respectively. The right axis is the calculated interaction energy $\Delta$ stemming from the count rate. The dashed line is the expected interaction energy for the twisted bilayer graphene, estimated from the enhancement in \textit{SaS} processes relative to diamond obtained in Ref.~\onlinecite{Parra-Murillo2016Stokesanti-StokesFunction}. (b) $I_{SaS}^{corr}(\varepsilon_{S,aS} = \pm$2070{\invcm}) as a function of the relative Raman cross-section above $\varepsilon_{S,aS}$, for different hydro-carbons and water, all measured with $P_L \sim 30$\,mW. The relative Raman cross-sections are estimated from the squared Raman peak area, $A_{Raman}^2$. All matrix elements and correlated {\it SaS} rates are taken as compared to the highest measured matrix element (cyclo-hexane).}
\label{figure3}
\end{figure}

Interestingly, the demonstration of momentum and energy conservation in this light-by-light scattering process is straightforward for each correlated {\it SaS} pair \cite{Saraiva2017PhotonicPairs}, while electronic Cooper pairs exist as a collective state inside superconductors, defying any attempts to address each pair individually. On the other hand, photon-photon interactions mediated by vacuum fluctuations are notoriously faint -- for instance, in the Atlas experiments, such interactions are observable, but only under very special conditions~\cite{atlas2017evidence}, in the really very high energies regime. As a result, the number of observed correlated {\it SaS} is extremely small, approximately one pair for every 10$^{15}$ incident photons. We here observe a rate of approximately 20 correlated {\it SaS} per second for Raman shifts below 1300{\invcm} in Fig.~\ref{figure1}(b). This rate is proportional to the interaction energy, which is the main energy scale that will determine if other analogous effects related to superconductivity will be observable. We estimate the interaction energy for photons scattered by diamond phonons at a Raman shift of $\varepsilon_{S,aS} = \pm$900{\invcm} from the transition probability $p=|\Delta|^2 dt^2/\hbar^2$ \cite{suppinfo}, where  $\Delta=V_0 |\alpha_L|^2$ is the transition amplitude \cite{Saraiva2017PhotonicPairs}. We conclude that $\Delta\gtrsim 10\,\mu$eV and then estimate an average attractive interaction energy $V_0\approx10$\,feV for diamond under our experimental conditions. 

Considering the dependence of the interaction $\Delta$ with the laser power ($P_L = |\alpha_L|^2 \hbar \omega_L R_L$), in Fig.~\ref{figure3}(a) we estimate how large this interaction strength may become if a more intense laser is used. The rate of pair production (filled circles) is proportional to the squared laser power ($P_L^2$), but with the absolute value depending on whether the frequency shift is below, at or above the phonon resonance (see open circles measured at two different $P_L$ values). Another parameter that may be explored in order to enhance $\Delta$ is the efficiency of the Raman scattering $M_q$. The intensity of pairs should, therefore, be also proportional to the squared Raman peak area $A_{Raman}^2$. We confirm this relationship by plotting $I_{SaS}^{corr}(\varepsilon_{S,aS} = \pm 2070{\invcm})$ as a function of the experimentally obtained $A_{Raman}^2$ above $\varepsilon_{S,aS}$ in different hydrocarbons and water (see Fig.~\ref{figure3}(b)). The listed materials are chosen here because they all exhibit a Raman peak near 2900{\invcm} (C-H and O-H vibrations) and no other Raman scattering contribution down to $\sim 2070${\invcm}. The observation of $I_{SaS}(\varepsilon_{S,aS}) \propto A_{Raman}^2$ in different materials is an ultimate proof that phonons are indeed responsible for the photon-photon scattering.

For completeness, we have measured the $\varepsilon_{S,aS}$ dependence of $I_{SaS}^{corr}(\varepsilon_{S,aS})$ for one liquid (decane, not shown) and, consistently, we could not observe correlated {\it SaS} above the highest frequency Raman mode at $\sim 2900${\invcm}. Therefore, the $I_{SaS}^{corr}(\varepsilon_{S,aS})$ asymmetry with respect to the phonon energy holds for both solids and liquids, indicating the universality of the correlated {\it SaS} phenomenon. The fact that virtually any transparent medium will generate pairs suggests that the photon pairs may be tailored in all its properties, such as energy, polarization, momentum and phase, by suitable choices of materials. Moreover, the input light source may be of any kind, as long as it is strong enough to actually generate pairs, and the consequences of the supercurrent analogy will serve as basis for new application proposals. The simplest of these consequences is the iris experiment (Fig.\ref{figure2}), which shows the \textit{S} and \textit{aS} photons crossing the material without the spread usually observed in light-phonon scattering. Similarly, there could be no spread in propagation time. Like in electronic superconductivity, these entanglement-derived properties should be a source of photonic state stability.

In our diamond experiment, typical orders of magnitude for the laser energy $E_L$, real phonon energies $E_q$, and transition amplitude $\Delta$ are 1\,eV, 0.1\,eV and 10$^{-5}$\,eV, respectively. Shen {\it et al.} \cite{Shen2002PCP} identify the phonon energy as the superconducting gap. We speculate that if a transition amplitude $\Delta$ reaches the phonon $E_q$ or the photon $E_L$ energies, new physical phenomena may happen. Specifically, the relation between the formation of the pairs and the vibration of the material points in the direction of using this technique to explore the material's properties beyond the information provided by standard intensity measurements. A more radical rupture would be the observation of speed of light renormalization, lending photons some finite mass, which would be central to the prediction of what collective bosonic condensate state (photonic liquid) might emerge. 

More efficient Raman scattering is needed in order to explore the meaning and usefulness of $\Delta$. For instance, a coincidence rate increase by a factor of 390 for resonant \textit{SaS} pairs was obtained in twisted bilayer graphene by engineering van Hove singularities \cite{Jorio2014Optical-PhononGraphene}. This may lead to interactions of the order of meV, as shown by the dashed line in Fig.~\ref{figure3}(a). Clearly, other experimental studies and a microscopic theory are needed for further advances.

We acknowledge L. M. Malard for helpful discussions. Financial support: CNPq (552124/2011-7, 307481/2013-1, 304869/2014-7, 460045/2014-8, 305384/2015-5, 309861/2015-2, FINEP(01.13.0330.00), CAPES (RELAI) and FAPERJ (E-26/202.915/2015, E-05/2016tTXE-05/2016). Correspondence should be addressed to A.J. (adojorio@fisica.ufmg.br).


%

\end{document}